\documentclass{cup}

\usepackage{latexsym}
\usepackage{graphicx}
\usepackage{multicol,multirow}
\usepackage{amsmath,amssymb,amsfonts}
\usepackage{mathrsfs}
\usepackage{amsthm}
\usepackage{rotating}
\usepackage{appendix}
\usepackage[authoryear]{natbib}
\usepackage{ifpdf}
\usepackage[T1]{fontenc}
\usepackage{times}
\usepackage{newtxmath}
\usepackage{textcomp}%
\usepackage{xcolor}%
\usepackage{hyperref}
\usepackage{lipsum}
\usepackage{subcaption}
\usepackage{tcolorbox}
\usepackage{float}
\newfloat{mybox}{tbp}{ext}
\restylefloat*{mybox}
\floatname{mybox}{Box}

\jname{Environmental Data Science}
\jyear{2020}
\jvol{4}
\jdoi{10.1017/eds.2020.xx}
\begin{document}

\begin{Frontmatter}

\title[Ethical AI for Environmental Sciences]{The Need for Ethical, Responsible, and Trustworthy Artificial Intelligence for Environmental Sciences}

\author*[1]{Amy McGovern}\orcid{0000-0001-6675-7119}\email{amcgovern@ou.edu}
\author[2,3]{Imme Ebert-Uphoff}\orcid{0000-0001-6470-1947}
\author[4]{David John Gagne II}\orcid{0000-0002-0469-2740}
\author[5]{Ann Bostrom}\orcid{0000-0002-6399-3404}

\address*[1]{\orgdiv{School of Computer Science and School of Meteorology}, \orgname{University of Oklahoma}, \orgaddress{\city{Norman}, \postcode{73019}, \state{OK},  \country{USA}}}

\address[2]{\orgdiv{Electrical and Computer Engineering}, \orgname{Colorado State University}, \orgaddress{\city{Fort Collins}, \postcode{80523}, \state{CO},  \country{USA}}}

\address[3]{\orgdiv{Cooperative Institute for Research in the Atmosphere}, \orgname{Colorado State University}, \orgaddress{\city{Fort Collins}, \postcode{80523}, \state{CO},  \country{USA}}}

\address[4]{\orgdiv{Computational and Information Systems Laboratory}, \orgname{National Center for Atmospheric Research}, \orgaddress{\city{Boulder}, \postcode{80301}, \state{CO},  \country{USA}}}

\address[5]{\orgdiv{Evans School of Public Policy \& Governance}, \orgname{University of Washington}, \orgaddress{\city{Seattle}, \postcode{98195}, \state{WA},  \country{USA}}}

\received{15 December 2021}
\revised{}
\accepted{}

\authormark{McGovern et al.}


\abstract{Given the growing use of Artificial Intelligence (AI) and machine learning (ML) methods across all aspects of environmental sciences, it is imperative that we initiate a discussion about the ethical and responsible use of AI.  In fact, much can be learned from other domains where AI was introduced, often with the best of  intentions, yet often led to unintended societal consequences, such as hard coding racial bias in the criminal justice system or increasing economic inequality through the financial system.  A common misconception is that the environmental sciences are immune to such unintended consequences when AI is being used, as most data come from observations, and AI algorithms are based on mathematical formulas, which are often seen as objective. 
In this article, we argue the opposite can be the case.  Using specific examples, we demonstrate many ways in which the use of AI can introduce similar consequences in the environmental sciences. This article will stimulate discussion and research efforts in this direction. As a community, we should avoid repeating any foreseeable mistakes made in other domains through the introduction of AI.  In fact, with proper precautions, AI can be a great tool to help {\it reduce} climate and environmental injustice.  We primarily focus on weather and climate examples but the conclusions apply broadly across the environmental sciences.}

\policy{This position paper discusses the need for the environmental sciences community to ensure that they are developing and using artificial intelligence (AI) methods in an ethical and responsible manner.  This paper is written at a general level, meant for the broad environmental sciences and earth sciences community, as the use of AI methods continues to grow rapidly within this community.}

\end{Frontmatter}

%%%%%%%%%%%%%%%%%%%%%%%%%%%%%%%%%%%%%%%%%%%%%%%%%%%%%%%%%%%%%%%%%%
\section{Motivation}

Artificial Intelligence (AI) and Machine Learning (ML) have recently exploded in popularity for a wide variety of environmental science applications \citep[e.g.][]{Reichstein2019,McGovern2019_bams,Gensini2021,Lagerquist2021,Hill2021,Schumacher2021, Gagne2020-kj}. Like other fields, environmental scientists are seeking to use AI/ML to build a linkage from raw data, such as satellite imagery and climate models, to actionable decisions. 

While the increase in applications of AI can bring improved predictions, e.g., for a variety of high-impact events, it is also possible for AI to unintentionally do more harm than good if it is not developed and applied in an ethical and responsible manner.  This has been demonstrated in a number of high-profile cases in the news outside of weather or climate \citep[e.g.,][]{tan2020film, WeaponsOfMath,RaceAfterTechnology}.  We argue that the potential for similar issues exists in environmental science applications and demonstrate how AI/ML methods could go wrong for these application areas.

On the other hand, AI can also be helpful for environmental sustainability.  AI is already being applied to enable the automated monitoring of our ecosystems to support accountability for climate justice. Applications include monitoring land cover changes to detect deforestation \citep{Karpatne2016-ti, Mithal2018-gp}, counting populations of endangered species in very-high resolution satellite data \citep{Duporge2021-cb}, and tracking bird populations in radar data \citep{Lin2019-xd, Chilson2019-uv}. Automated monitoring of retrospective and real-time datasets can help interested parties to monitor environmental trends and respond appropriately. 

There are both instrumental and consequential as well as principled ethical  perspectives such as duties and virtues for AI environmental scientists to consider. These are entangled in the biases and pitfalls this paper explores.  Debiasing has the potential to address both. 

Note, we assume the reader is generally familiar with the concepts of AI and ML and do not focus on any specific AI/ML methods but instead on the applications across environmental sciences.  For brevity, we refer to AI/ML as AI throughout the rest of the paper.

%%%%%%%%%%%%%%%%%%%%%%%%%%%%%%%%%%%%%%%%%%%%%%%%%%%%%%%%%%%%%%%%%%
\section{How AI can go wrong for environmental sciences}

Although Arthur C. Clarke's famous line "Any sufficiently advanced technology is indistinguishable from magic" was not directed at AI, it may seem that way to many new users of AI.  It seems as if one simply has to give an AI method data and use a pre-made package and an amazing predictive model results.  The problem with this approach is that AI is not magic and can be led astray in a number of ways, which we outline here.   

Box \ref{fig:ai_issues} provides a non-exhaustive list of ways in which AI can go wrong for environmental sciences, and other issues that can arise.  We discuss this list in more depth below and provide examples from the environmental sciences to illustrate them.

\begin{mybox}[h!]
\begin{tcolorbox}[title=Ways in which AI can go wrong for environmental sciences,center title,colback=blue!10!white,colbacktitle=blue!50!red,fonttitle=\bfseries]

\textbf{Issues related to training data:}
\begin{enumerate}
    \item Non-representative training data, including lack of geo-diversity 
    \item Training labels are biased or faulty
    \item Data is affected by adversaries
\end{enumerate}
\textbf{Issues related to AI models:}
\begin{enumerate}
    \item Model training choices    
    \item Algorithm learns faulty strategies 
    \item AI learns to fake something plausible 
    \item AI model used in inappropriate situations
    \item Non-trustworthy AI model deployed
    \item Lack of robustness in the AI model
\end{enumerate}
\textbf{Other issues related to workforce and society:}
\begin{enumerate}
    \item 
    Globally applicable AI approaches may stymie burgeoning efforts in developing countries.
    \item Lack of input or consent on data collection and model training
    \item
    Scientists might feel disenfranchised. 
    \item
    Increase of CO$_2$ emissions due to computing
\end{enumerate}

\end{tcolorbox}
\caption{A non-exhaustive list of issues that can arise though the use of AI for environmental science applications. \label{fig:ai_issues}}
\end{mybox}

%%%%%%%%%%%%%%%%%%%%%%%%%%%%%%%%%%%%%%%%%%%%%%%%%%%%%%%
\subsection{Issues related to training data}
\label{sec:data-issues}

AI models follow the well-known computer science adage:  Garbage in, garbage out.  If the training data is not representative of what we actually want to use the AI model for, then it will produce models that do not accomplish our goals.  Because training data are key, issues related to them are particularly common and hard to avoid.  First of all, as we demonstrate in the examples below, it is extremely difficult to create a data set that does not have some kind of bias or other shortcoming, even if it was developed with great care and with the best intentions. Secondly, while the data developers might be aware of many (but usually not all) of the shortcomings, the scientist {\it using} the data to train AI models might be completely unaware of them, since there is no standard (yet!) in the community to document these shortcomings and include them with the data. Furthermore, whether a data set is biased also depends on the application.  It might be perfectly fine to use a data set for one purpose, but not for another.  Lastly, there is no established set of tests to check for the most common biases in environmental science data sets. 

It is important to understand that when AI models are trained on biased data, they {\it inherit} those biases.  This phenomenon is known as {\it coded bias} \citep{WeaponsOfMath,tan2020film} and is easy to understand - if a hail data set shows hail occurring only in populated areas (see Example 1(a) below), the AI system trained on this data set is likely to also predict hail only in highly populated areas, making the AI model just as biased to population density as the data it was trained on.  Since established algorithms are often used in regions or under circumstances other than the origin of the data, such an algorithm can even perpetuate the bias beyond the regions and circumstances it was trained on.

%%%%%%%%%%%%%%%%%%%%%%%%%%%
\subsubsection{Non-representative training data}
\label{sec:non-representative}

\textbf{Statistical distribution of training, testing, and validation data:} 
If the training data is non-representative statistically, the output is likely to be biased.  This could include geographic or population biases (discussed in more detail in Section \ref{sec:labels}) but it could also include unintentional temporal biases.  For example, if an AI model is trained on data from 1900-1970, 1970-2000 is used as a validation set, and then 2000-2020 is used as a test set, the model may not be properly accounting for climate change.  
Sensor limitations provide other examples.  Many sensors require sunlight for high quality observations, and thus many phenomena are under-observed at night.  Similarly, inferring certain surface conditions from satellites using passive sensors tends to be easier for clear sky conditions.

\begin{figure}
    \centering
    \includegraphics[width=5in]{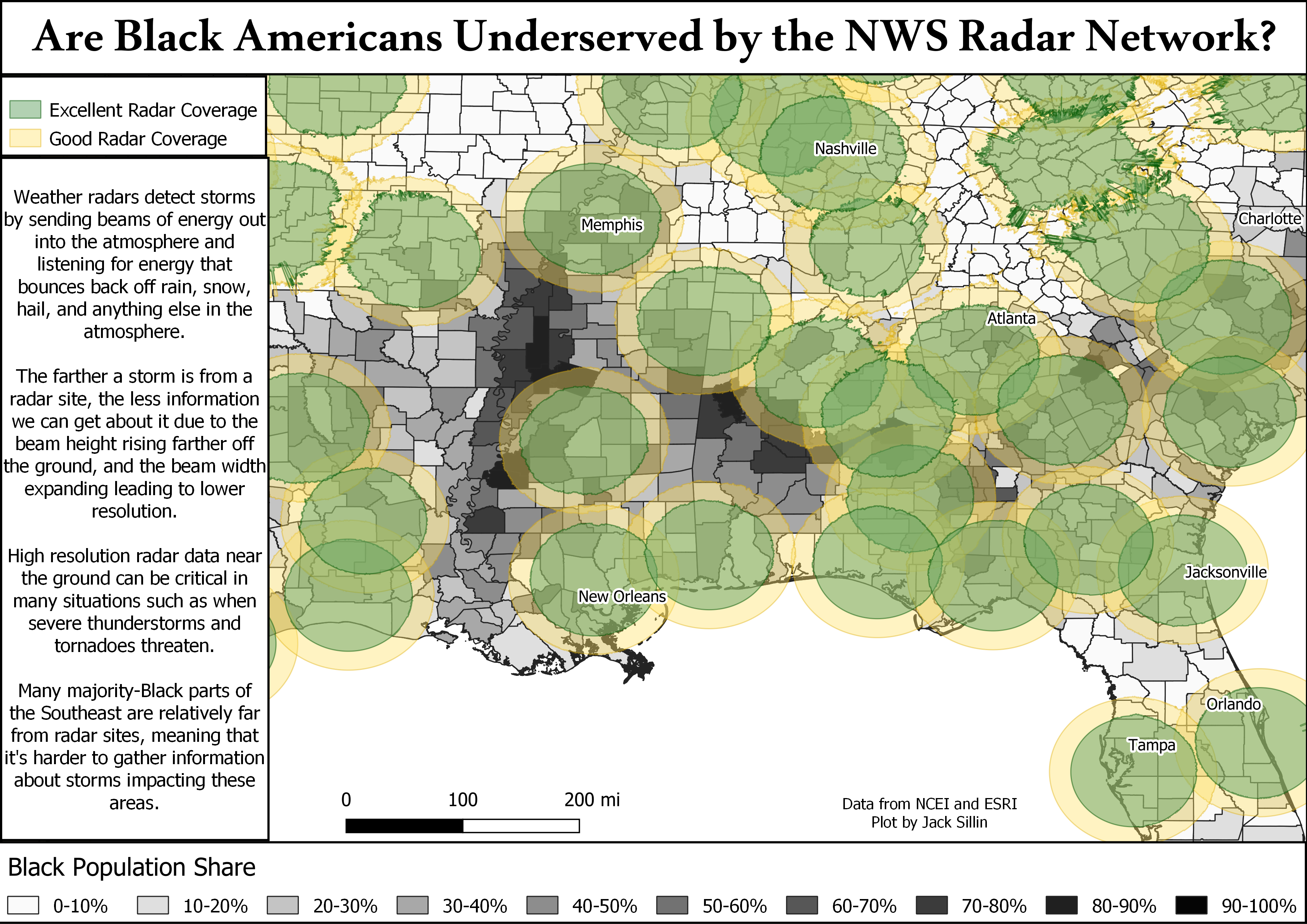}
    \caption{Coverage of the national Doppler weather network (green and yellow circles) overlaid with the black population in the southeast United States, courtesy of Jack Sillin. This is an example of non-representative data (Section \ref{sec:non-representative})}
    \label{fig:southeast_radar}
\end{figure}

\noindent
\textbf{Lack of geo-diversity:}  
In order to enable AI algorithms to facilitate environmental justice we thus need to ensure that different populations are well represented in their training data.  This includes ensuring that the data is diverse geographically, including addressing population biases and biases in sensor placements that could affect the ability to obtain a statistically representative training set.  For example, the national radar network has coverage gaps which can inadvertently under-represent some populations, as indicated in Figure \ref{fig:southeast_radar} \citep{Sillin2021,ShepherdRadar2021}. We should explicitly take population representation into account for the design of future sensor networks and seek to close gaps in existing sensor networks by placing additional sensors in strategic locations. Furthermore, we can put AI to good use by developing AI algorithms to estimate sensor values at the missing locations based on other sensor types, such as generating synthetic radar imagery from geostationary satellite imagery \citep{hilburn2021development}.  

%%%%%%%%%%%%%%%%%%%%%%%%%%%%%%%%%%%%%%%%%%%%%%%%%%%%%%%%%%%%%%%%%%
\subsubsection{Training labels are biased or faulty}
\label{sec:labels}

\begin{figure}
    \centering
    \begin{subfigure}{0.45\textwidth}
        \includegraphics[width=\textwidth]{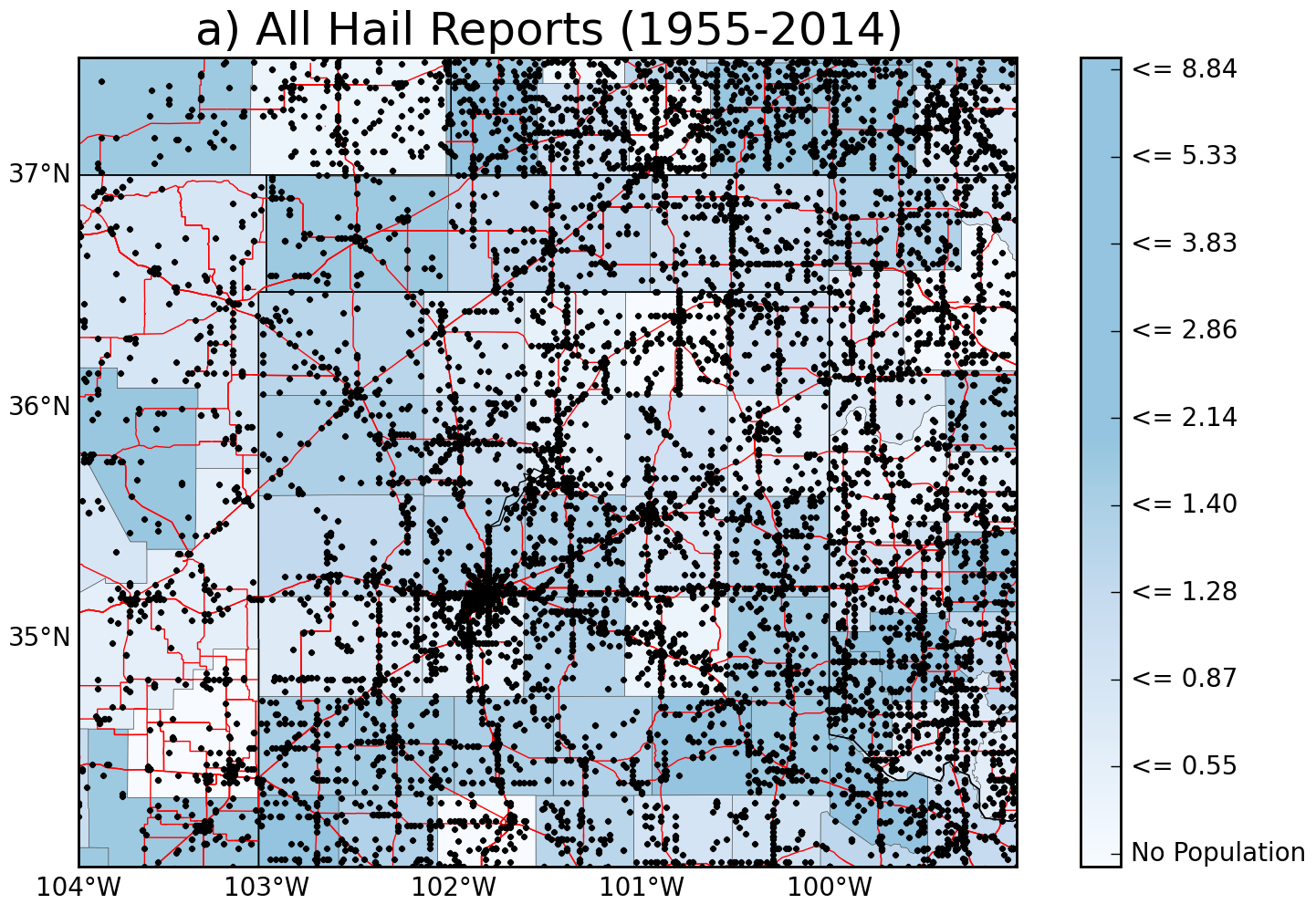}
        \caption{Hail reports follow population}
        \label{fig:hail_population}
    \end{subfigure}
    \hfill
    \begin{subfigure}{0.45\textwidth}
        \includegraphics[width=\textwidth]{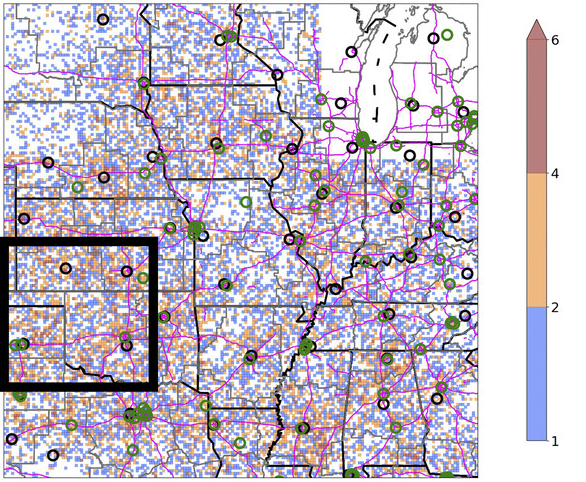}
        \caption{Tornado reports follow population}
        \label{fig:tornado_population}
    \end{subfigure}
    \caption{Hail and tornado reports both show a clear population bias with reports occurring more frequently along roads and cities.  This can be seen as examples of (i) data being missing in low population areas (Section \ref{sec:non-representative}) and (ii) faulty labels (Section \ref{sec:labels}).  Panel a is from \citet{AllenHail2015}; Panel b is from \citet{Potvin2019tornadoreports}}
    \label{fig:population_bias}
\end{figure}

\textbf{Human generated data sets:}
Obtaining data for environmental science prediction tasks is often quite challenging, both in terms of finding the data and in terms of identifying a complete data set.  For example, if one was to predict hail or tornadoes but only to use human reports, there is a known bias towards areas with higher population \citep{AllenHail2015, Potvin2019tornadoreports} as shown in Figure \ref{fig:population_bias}). If more people live in an area, there is a higher chance that someone observes and reports a hail or tornado event. 
This might bias the AI model to over-predict urban hail/tornadoes and under-predict rural hail/tornadoes. 

A related issue is that even if a diverse geographical area is represented in the dataset, the AI model will focus on patterns that are associated with the most common regimes and areas where the hazard occurs most frequently. It also may ignore patterns that affect a small portion of the dataset since the contribution to the overall training error is relatively small. For example, solar irradiance forecasts from an AI model trained across multiple sites in Oklahoma had higher errors at the sites in the Panhandle compared with central Oklahoma due to limited representation of Panhandle weather in the dataset \citep{Gagne2017-vw}.

\begin{figure}
    \centering
    \begin{subfigure}{0.45\textwidth}
        \includegraphics[width=\textwidth]{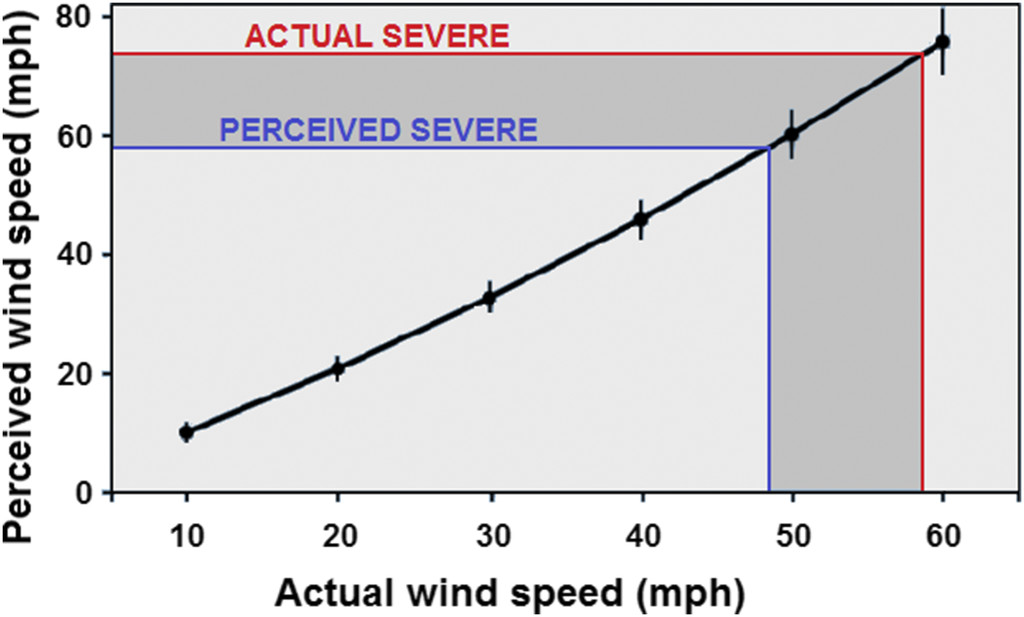}
        \caption{Human estimates and observations of wind speed}
        \label{fig:wind_speeda}
    \end{subfigure}
    \hfill
    \begin{subfigure}{0.45\textwidth}
        \includegraphics[width=\textwidth]{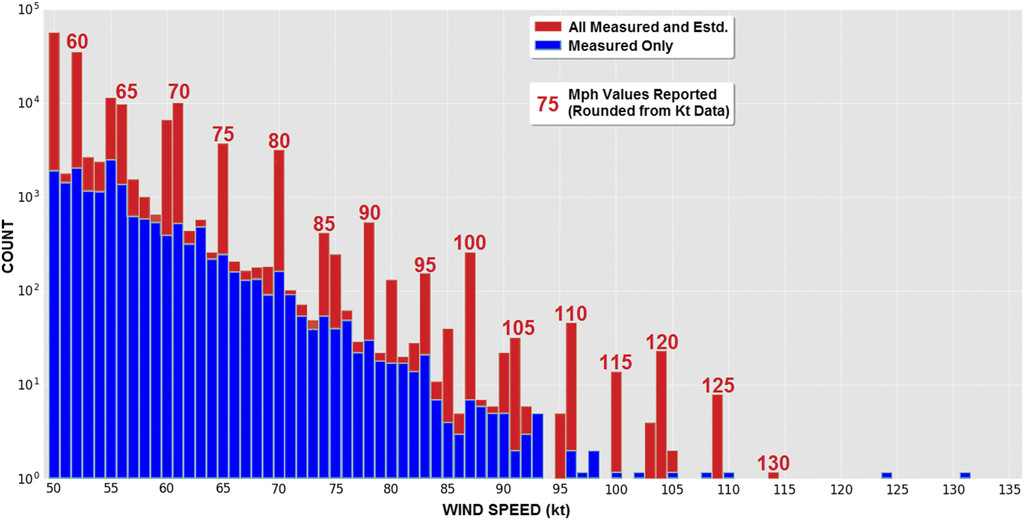}
        \caption{Human reports of max wind speed versus measured max wind speed from \citet{EdwardsWindReliability:2018}.}
        \label{fig:wind_speedb}
    \end{subfigure}
    \caption{Human reports of wind speed show biases in both perception of the speed itself (Panel a) and in the binning of the data (Panel b).  Both panels are from \citet{EdwardsWindReliability:2018}. This example highlights both non-represenative training data (Section \ref{sec:non-representative})
    and
    biased and faulty labels (Section \ref{sec:labels}) 
    }
    \label{fig:wind_speed}
\end{figure}

Figure \ref{fig:wind_speed} illustrates the typical biases of humans to estimate wind speed due to two effects. Figure \ref{fig:wind_speeda} shows that humans tend to overestimate wind speed, and thus tend to classify it as severe at levels that are not actually considered severe.  Thus, human reported wind speed results in training data with significant bias toward higher wind speeds.
Figure \ref{fig:wind_speedb} illustrates that humans tend to assign wind speeds in discrete increments, with a strong bias toward multiples of five, i.e., 60, 65, 70, etc., kt.  This example also demonstrates how  non-representative training data and the biases could result in an AI algorithm that does not predict true wind speed impacts correctly.

\noindent
\textbf{Sensor generated data sets:}
A potential response to this data issue could be to use only sensor-based data, such as radar-derived hail estimates.  We often assume that sensors are inherently objective yet that is not guaranteed.  First, the data from these sensors must be interpreted to create a full data set.  For radar-derived hail data, there are well known overprediction biases to the data \citep[e.g.,][]{MurilloMesh2019}. 
Second, data from sensors may also be incomplete, either due to sensor errors or missing sensors.  With many sensors, there are areas which are well represented and areas where data is lacking.  Sensor placement often depends on geological features, e.g., there might be a lack of sensors at remote mountain tops or in empty fields.  
Economical reasons can also come into play in the placement of sensors, such as the placement of crowd-sourced sensors where there are more affluent people. Consider for example the deployment of air quality sensors.  While the EPA has a network of just over 1,100 air quality sensors that are relatively evenly distributed throughout the US, this is not the case for the popular new generation of low-cost air quality sensors, called PurpleAir.  There are now over 5,500 PurpleAir sensors deployed in the US, but they are deployed in significantly Whiter, higher income areas than the national average
\citep{desouza2021distribution}.  Access to these sensors in less-privileged communities is needed to democratize air pollution data \citep{desouza2021distribution}.

%%%%%%%%%%%%%%%%%%%%%%%%%%%%%%%%%%%%%%%%%%%%%%%%%%%%%%%%%%%%%%%%%%
\subsubsection{Data is affected by adversaries}
\label{sec:adversarial-data}

Adversarial data is a well-known problem in machine learning \citep[e.g.,][]{Adversarial:models,GoodfellowEtAl:MakeMLRobust,ADV:Diochnosetal:neurips2018}. Adversaries can cause an AI model to either learn a faulty model or to be used incorrectly when applied to real-world data.  For example, there are well-known examples of minor changes to speed signs causing a computer vision system to see a 35 mph speed limit as an 85 mph speed limit, something that could be quite dangerous in an autonomous car.  

AI applied to environmental science data has to contend with two kinds of adversaries:  humans and the environment.  One needs to be aware of both types of adversaries when training and applying the AI models to the real-world.

\begin{figure}
    \centering
    \begin{subfigure}{0.25\textwidth}
        \includegraphics[width=\textwidth]{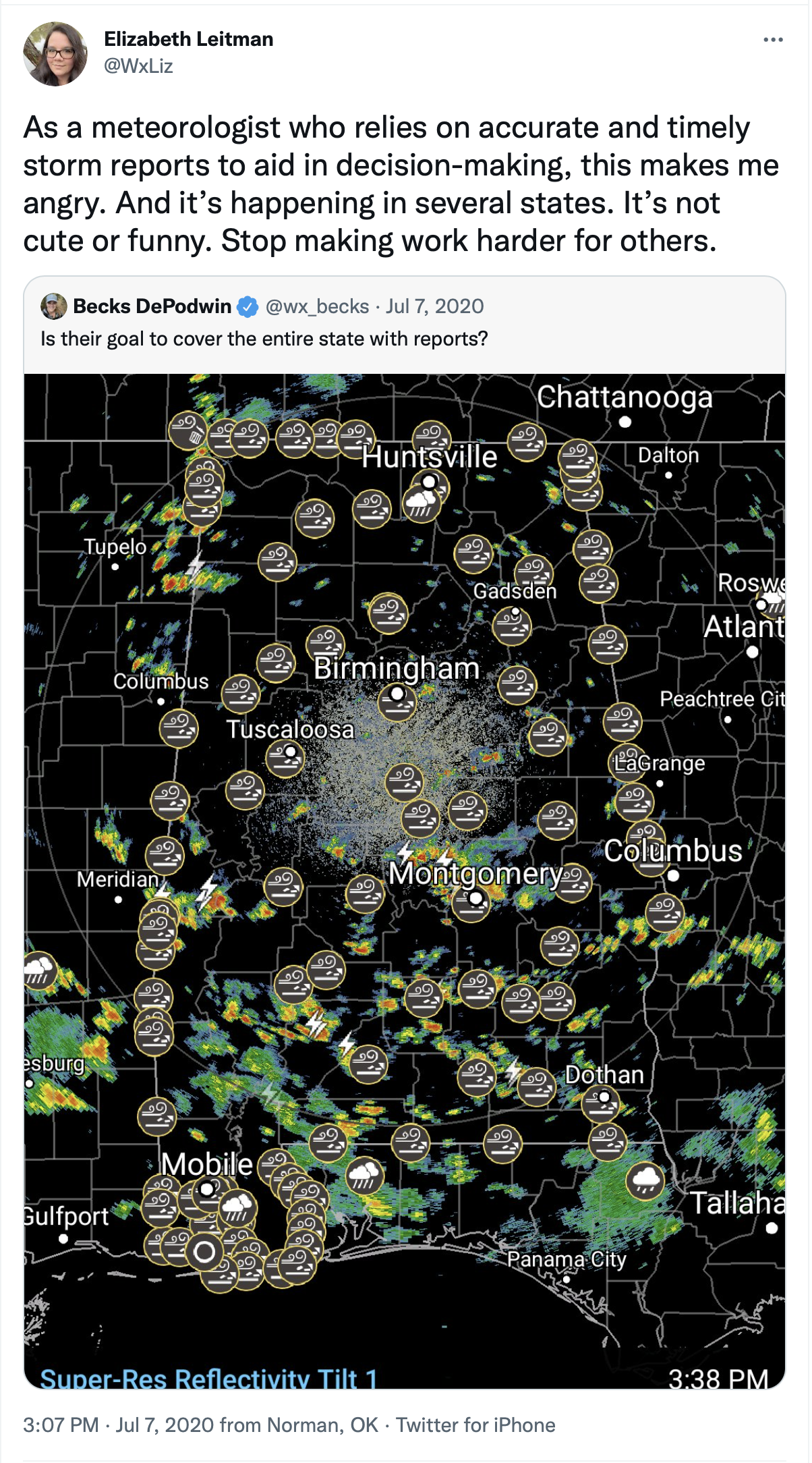}
        \caption{Screenshot of hacked data from mPing from \citet{MpingWaPo}}
        \label{fig:mping}
    \end{subfigure}
    \hfill
    \begin{subfigure}{0.7\textwidth}
        \includegraphics[width=\textwidth]{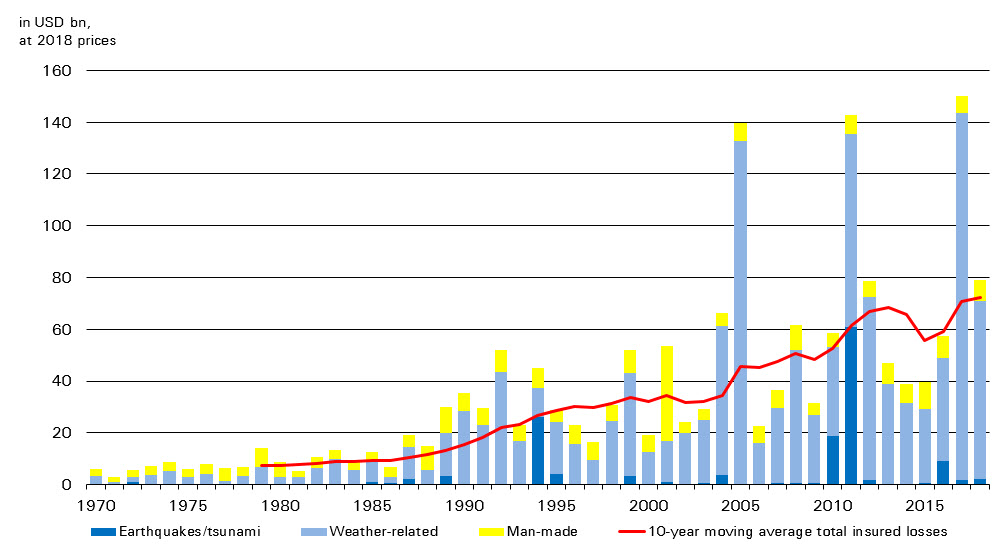}
        \caption{Catastrophe related insured losses from 1970 to 2018 from \citep{InsuredLosses2018,WorldEconomicInsurance}.  With at least 10\% of weather-based insurance claims estimated as fraudulent \citep{InsuranceFraud}, there is increasing motivation for humans to report incorrect severe weather data}
        \label{fig:insurance_loss}
    \end{subfigure}
    \caption{Humans sometimes create adversarial data, which may be ingested by an AI model (Section \ref{sec:adversarial-data}).  While in example (a) the user intent is to create false inputs, in example (b) the motivation for the false reports is more for personal financial gain (insurance fraud).  Nevertheless, both can cause problems for AI by directly affecting reports databases used by AI models} 
    \label{fig:human_adversary}
\end{figure}

\noindent
\textbf{Human adversaries: } When using human reported data there is always the possibility of users intentionally adding bad data.  Figure \ref{fig:human_adversary} provides two examples.  Figure \ref{fig:mping} provides the screenshot of wind reported by someone who hacked the crowd-sourced mPing system for the state of Alabama.  Note the long line of wind reported outlining the state border which clearly represents an intentional input of incorrect information.  Such data would be detrimental to any AI model trained on this data. A second source of human adversarial data is insurance fraud \citep{Davila2005-xk}.  As the number of expensive natural disasters are increasing (Figure \ref{fig:insurance_loss}), the number of fraudulent weather-related insurance claims is increasing, with estimates of fraudulent reports hovering around 10\% \citep{InsuranceFraud}. Such fraudulent reports can also affect training data for AI by appearing in the databases of severe weather reports.

\begin{figure}
    \centering
    \begin{subfigure}{0.45\textwidth}
        \includegraphics[width=\textwidth]{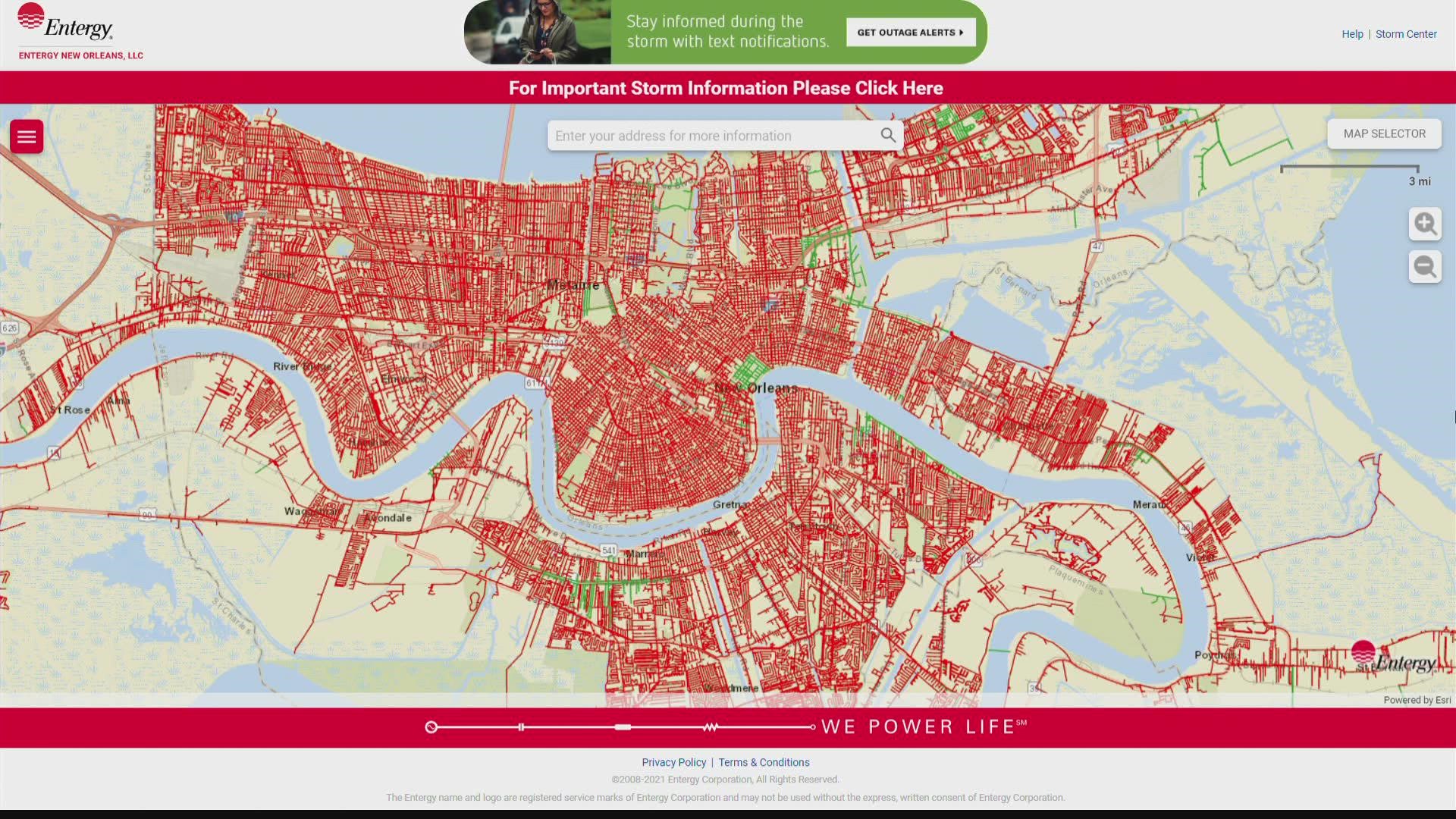}
        \caption{Power outages following a hurricane}
        \label{fig:power_outages}
    \end{subfigure}
    \hfill
    \begin{subfigure}{0.45\textwidth}
        \includegraphics[width=\textwidth]{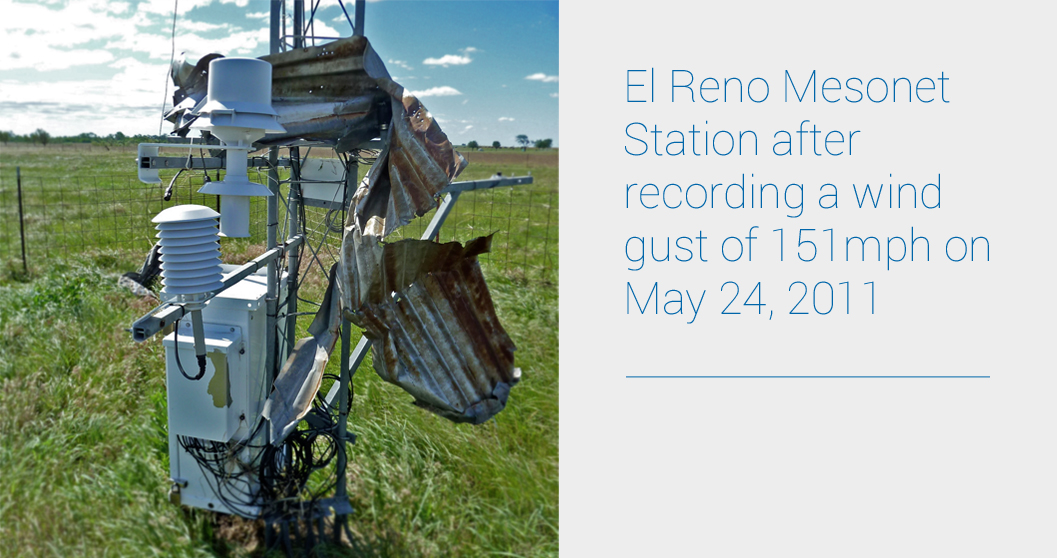}
        \caption{A destroyed Oklahoma Mesonet station following a severe wind gust}
        \label{fig:el_reno}
    \end{subfigure}
    \caption{Weather also creates its own adversaries (Section \ref{sec:adversarial-data}) by creating power outages (a) or destroying sensors (b), especially during major events.  Graphics from  \citet{HurricaneOutages,MesonetTop20}}
    \label{fig:weather_adversary}
\end{figure}

\noindent
\textbf{Weather as an adversary:} Weather can act as its own adversary, especially when collecting observations for extreme events.  For example, when seeking to observe the severity of storms, sensors can fail due to power outages caused by a strong storm (Fig.\ \ref{fig:power_outages}) or a strong storm can even destroy sensors (Fig.\ \ref{fig:el_reno}).

%%%%%%%%%%%%%%%%%%%%%%%%%%%%%%%%%%%%%%%%%%%%%%%%%%%%%%%
\subsection{Issues related to the AI model}

In addition to problems caused by bias or other complications in the training data, AI models can also develop issues on their own. Examples are given in this section.

%%%%%%%%%%%%%%%%%%%%%%%%%%%%%%%%%%%%%%%%%%%%%%%%%%%%%
\subsubsection{Model training choices}
Model training choices will affect every aspect of the AI model.  While AI methods are based on mathematical equations and thus often seen as ``objective", there are countless choices that a scientist has to make that greatly affect the results. These choices include:
%\begin{itemize}
%\item 
(i) which attributes (e.g., which environmental variables) to include in the model;
%item
(ii)  which data sources to use;
%\item
(iii) how to preprocess the data (e.g., normalizing the data, removing seasonality, applying dimension reduction methods);
%\item
(iv) which type of AI model to use (e.g., clustering, random forest, neural networks, etc.),
%\item
(v) how to choose hyper parameters (e.g., for random forest - how many trees, maximal depth, minimal leaf size, etc).
%\end{itemize}
Each of these choices has significant impact and can lead to vastly different results with severe consequences.
For example, the choice of spatial resolution for the output of a model can be crucial for environmental justice.  Training an AI model to predict urban heat at a low spatial resolution may average out, and thus overlook, extreme values in small neighborhoods, while using a higher spatial resolution could reveal those peaks but potentially introduce noise. 

%%%%%%%%%%%%%%%%%%%%%%%%%%%%%%%%%%%%%%%%%%%%%%%%%%%%%%%%%%%%%%%%%%
\subsubsection{Algorithm learns faulty strategies} 

AI models are tasked to learn patterns in the data that help them come up with good estimates or classifications.  Sometimes, even if the data are not faulty in the traditional sense, they may contain spurious correlations that are not representative of the real world and that the AI model learns to exploit. This issue could be discussed in the section on data-related issues (Section \ref{sec:data-issues}), but we chose to include it in this section since it is so closely coupled with AI model development, and often only diagnosed once an AI model has been developed.

\begin{figure}
    \centering
    \begin{subfigure}{0.45\textwidth}
        \includegraphics[width=\textwidth]{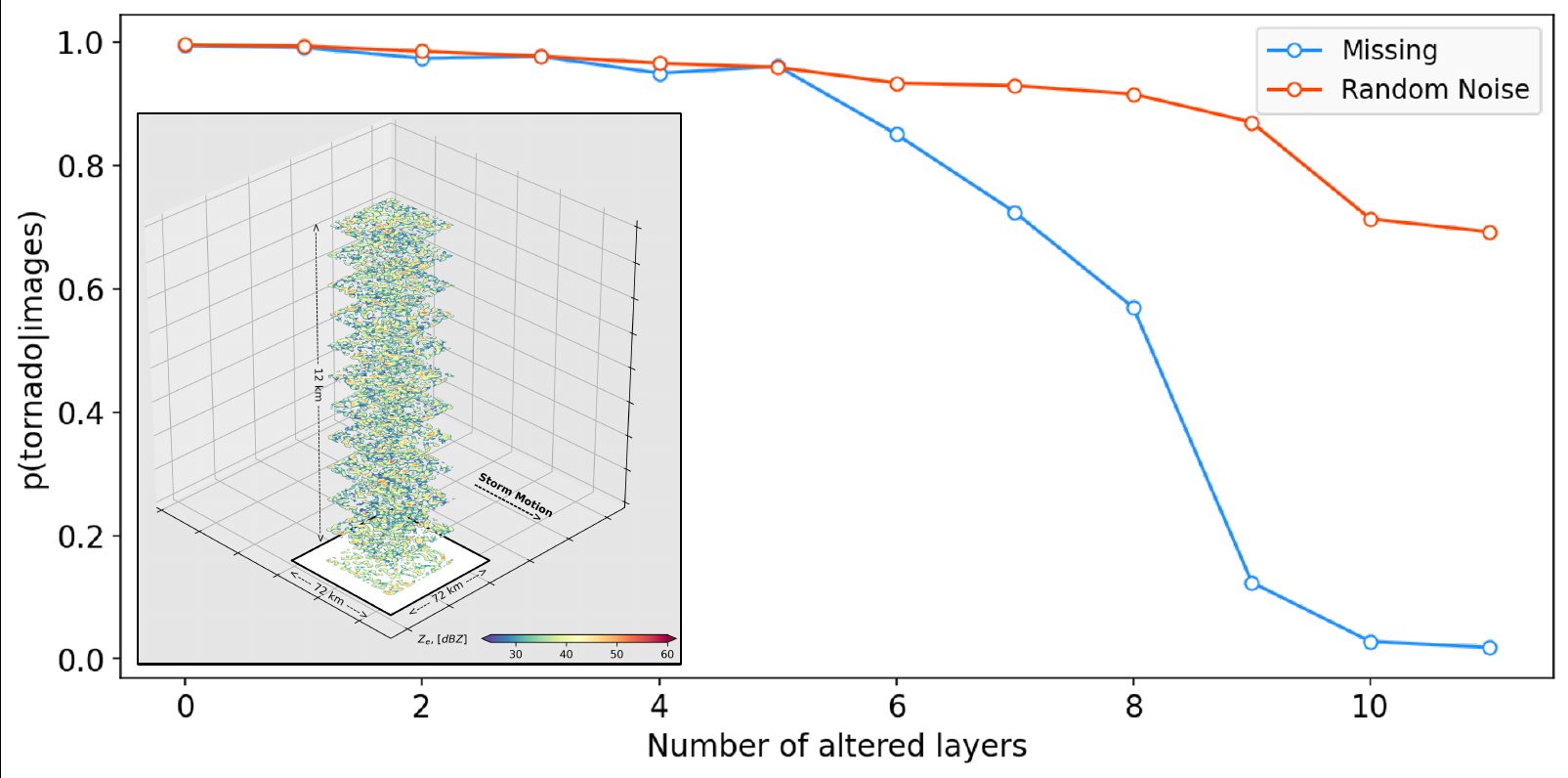}
        \caption{Tornado prediction model from \citet{Lagerquist2020} and Randy Chase \citep{chaseAMS2022}.  When examining the model, we discovered that adding random noise to the reflectivity did not decrease predictions as expected.}
        \label{fig:noise_panel}
    \end{subfigure}
    \hfill
    \begin{subfigure}{0.45\textwidth}
        \includegraphics[width=\textwidth]{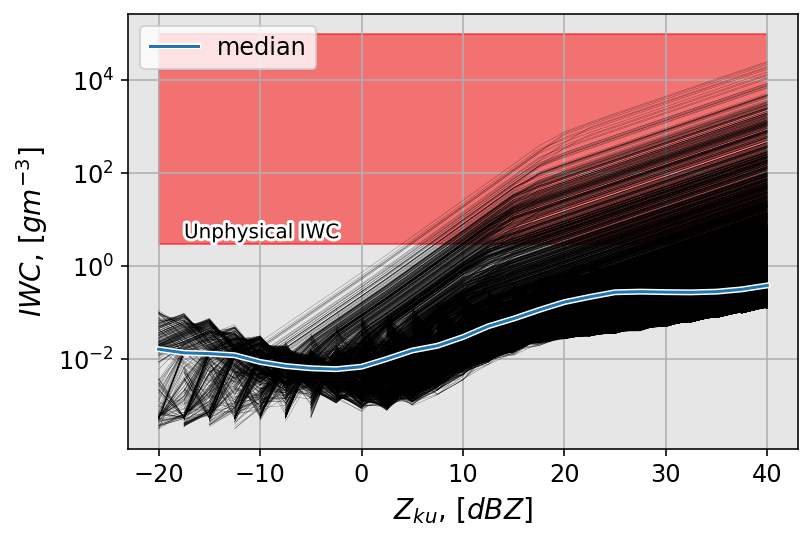}
        \caption{Example of a partial dependency plot \citep{Molnar2018} which shows how an AI model could produce unphysical results when given unexpected inputs (Section \ref{sec:model_applied_wrong}).  From \citep{ChaseRadar2021}}
        \label{fig:unphysical_model}
    \end{subfigure}
    \caption{Examples of an AI learning to fake something plausible (Section \ref{sec:AI_faking}) and of a model generating unexpected output when applied to data outside of the range of expected data (Section \ref{sec:model_applied_wrong})}
    \label{fig:incorrect_model}
\end{figure}

One approach to identifying when the AI has learned a faulty strategy is to either use interpretable or explainable AI (XAI) \citep[e.g.,][]{ras2018explanation,Molnar2018,XAI_book,McGovern2019_bams,ebert2020evaluation}.  In the case of interpretable AI, the models are intended to be understandable by human experts \citep{Rudin2019} while explainable AI attempts to peer inside the black box of more complicated models and identify the internal strategies used by the models.  Both approaches allow a human expert to examine the model and potentially to identify faulty strategies before deploying a model.

Figure \ref{fig:noise_panel} is an example of how an AI model can learn a faulty strategy and how this can be discovered through XAI.  In this example, a deep learning model was trained to predict tornadoes \citep{Lagerquist2020} and then examined to identify what the model had learned.  Unexpectedly, we discovered that the model was still predicting a high probability of tornadoes even if the reflectivity image contained mostly noise \citep{chaseAMS2022}.

%%%%%%%%%%%%%%%%%%%%%%%%%%%%%%%%%%%%%%%%%%%%%%%%%%%%%%%%%%%%%%%%%%
\subsubsection{AI learns to fake something plausible}
\label{sec:AI_faking}

The emergence of an advanced type of AI algorithm, namely the Generative Adversarial Network (GAN) introduced by \cite{goodfellow2014generative}, yields the ability to generate imagery that looks extremely realistic.  The result of a GAN can be seen as one possible solution (an ensemble member) which might not be representative of the distribution of possible solutions.  
%Nevertheless, \citet{stengel2020adversarial} demonstrate that GANs have the ability to generate imagery with better small scale physical properties than existing methods.
An important question is how forecasters would respond to GAN imagery with intricate small scale features that look overwhelmingly realistic, but might not be as accurate as they appear to be.
\citet{Ravuri2021-dg} conducted a cognitive assessment study with expert meteorologists to evaluate how forecasters respond to GAN imagery of radar for precipitation nowcasting. They concluded that for their application {\it forecasters made deliberate judgements of the predictions by relying on their expertise, rather than being swayed by realistic looking images} (Suppl.\ C.6. of \citet{Ravuri2021-dg}).
Nevertheless, the jury is still out whether that is generally the case.  Such methods might sometimes trick forecasters into greater trust than is warranted. 
Similar methods are used to generate deep fake images and videos \citep{IEEEDeepFakes} which have successfully fooled people into believing misinformation \citep{Vaccari2020-jh}.  If similar approaches are used in environmental science, trust in AI models could be lost.

One of the major dangers of generative models used in contexts where they are parameterizing fine scale data \citep{Gagne2020-kj} or filling in missing data \citep{Geiss2021-ob} is the lack of guarantees about what data they use to fill in the gaps. If an ML model is trained without any physical constraints, it is possible for the algorithm to insert spurious weather features. If this occurs in otherwise data sparse regions, forecasters may not have any additional data to discredit the false prediction for many hours. More subtle discrepancies would be the most problematic since they are harder for forecasters to catch. 

%%%%%%%%%%%%%%%%%%%%%%%%%%%%%%%%%%%%%%%%%%%%%%%%%%%%%%%%%%%%%%%%%%
\subsubsection{Models used in inappropriate situations}
\label{sec:model_applied_wrong}

Another model-based issue that can arise is when an AI model is trained for one situation, perhaps a specific weather regime or for a specific geographical area, and then applied to new situations where it is not physically meaningful.  In such situations, the AI may predict incorrectly, without giving any warning, which could lead to loss of life and environmental injustice.

For example, Figure \ref{fig:unphysical_model} shows a partial dependence plot \citep{Molnar2018} of the predictions of an AI method predicting total ice water content (i.e., mass of snow in a cloud) from radar data \citep{ChaseRadar2021}. If input given to the model is uncharacteristic of the training dataset, the output may be not physically realistic. This issue also comes up frequently when ML is used to replace Numerical Weather Prediction parameterization schemes. If the ML model is transferred to a different climate or geographic regime or even to the hemisphere opposite where it was trained (where the sign of many common variables are flipped), it often performs quite poorly.

%%%%%%%%%%%%%%%%%%%%%%%%%%%%%%%%%%%%%%%%%%%%%%%%%%%%%%%%%%%%%%%%%%
\subsubsection{Non-trustworthy models deployed}
    
\begin{figure}
    \centering
        \includegraphics[width=0.8\textwidth]{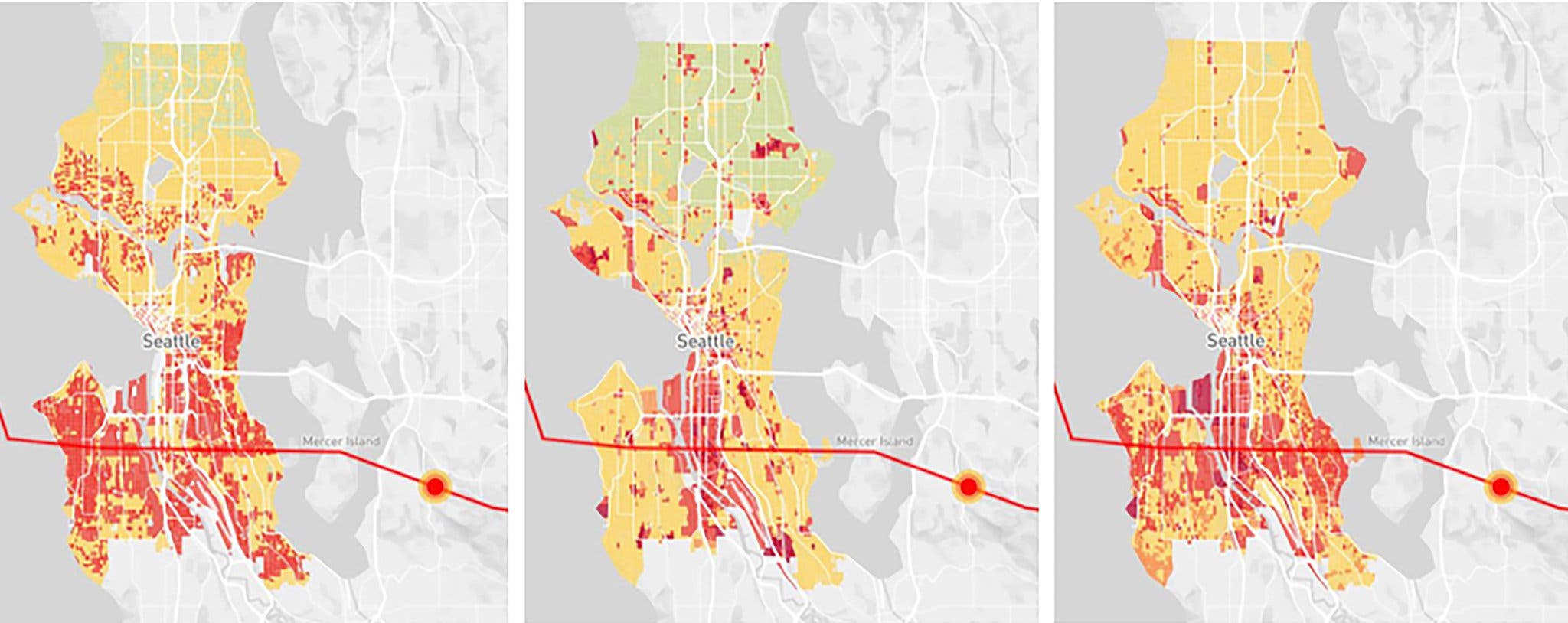}
        \caption{Three vastly different damage predictions for the same hypothetical 7.0-magnitude Seattle-area earthquake delivered by different versions of the same AI system. 
        Figure from \cite{fink2019high}} %\url{https://www.nytimes.com/2019/08/09/us/emergency-response-disaster-technology.html} 
        %and ask NYT / Seattle Office of Emergency Management for permission to use figure]
        \label{fig:earthquake}
\end{figure}

Ensuring that a model is trustworthy is a key part of ethical and responsible AI.  Often AI systems are {\it deployed} before they are ready, in particular before they have been thoroughly tested and validated by domain experts. 
Fig.\ \ref{fig:earthquake} shows the results of a commercial AI system predicting damage caused by a hypothetical 7.0-magnitude earthquake in the Seattle area. The three panels show results from three different version of the system delivered to the city of Seattle to assist them to plan where to establish earthquake shelters. The widely differing results in the three panels arose from using different data sources, but also from programming errors (e.g., counting each apartment in a high-rise building as a separate structure). 
Deploying the model before it was ready for operation  forced the city 
to revamp nearly completed plans for sheltering earthquake-displaced residents that were developed using the original version \cite{fink2019high}. The trust in the AI system was so eroded that the city considered terminating their contract, and only kept it because the company found an external funder and could offer it to Seattle for free. 

Typical mistakes to be avoided, illustrated by this example, are developing a system without engaging domain experts or local end-users, deploying a system before it has been rigorously validated, and over-promising the system's accuracy to the end user without any scientific evidence.  For the latter, there are many new methods for uncertainty quantification in machine learning models that we as a community should start utilizing routinely for this purpose \citep[e.g.,][]{orescanin2021bayesian,barnes2021adding}.

%%%%%%%%%%%%%%%%%%%%%%%%%%%%%%%%%%%%%%%%%%%%%%%%%%%%%%%%%%%%%%%%%%

%%%%%%%%%%%%%%%%%%%%%%%%%%%%%%%%%%%%%%%%%%%%%%%%%%%%%%%%%%%%%%%%%%
\subsubsection{Lack of robustness}

While a lack of robustness could also be categorized under models being inappropriately applied, adversarial data, or even a lack of trust, we separate it into its own category to highlight the need for robustness in ethical and responsible AI.  Robust AI for environmental sciences will fail gracefull when outputs are out of range (e.g. Section \ref{sec:model_applied_wrong}) and will also constrain the outputs by the laws of physics. For instance, when preparing a workshop tutorial, one of the authors was demonstrating a simple ML method that could predict a local temperature.  When presented with an input that was two standard deviations outside of its training data, the method suddenly predicted a temperature hotter than the surface of the sun.  Such predictions may contribute to a lack of trust and create environmental injustices.

%%%%%%%%%%%%%%%%%%%%%%%%%
\subsection{Other issues related to workforce and society}
\label{sec:other_issues}

We have discussed ways that both the training data and the AI model could go wrong.  Here we focus on the broader picture of how AI is applied and developed and how that relates to environmental justice. This aspect also needs to be addressed in developing ethical and responsible AI for environmental sciences. 

%%%%%%%%%%%%%%%%%%%%%%%%%%%%%%%%%%%%%%%%%%%%%%%%%%%%%%%%%%%%%%%%%%
\subsubsection{Globally applicable AI approaches may stymie burgeoning efforts in developing countries}
Large organizations developing and deploying AI systems, including government agencies and private companies, generally want their AI systems to be globally applicable. People in countries outside that of the provider organization can benefit from technological advances that may not otherwise be available to them. Provider organizations can benefit from a broader set of customers and stakeholders. These kinds of global activities work best when researchers and companies directly engage with different local populations and incorporate their needs and concerns into the design process.  For example, Google's crisis team has been working with multiple local and government partners in India and Bangladesh to create a hyper-local flood forecasting product \citep{GoogleFloodBlog2021}.  

However, the relative ease of deploying AI on global datasets can also increase the temptation for organizations to automate deployment without fully monitoring its impact. Without direct engagement with various populations, organizations may not be aware of the unintended consequences of their algorithms. They also may not have access to observations in these areas to validate their performance. They may also undercut local providers who are more aligned with the community but do not have the same resources and economies of scale to compete on price, especially if the service is offered for free or is supported by advertising revenue. Government meteorological centers in both developed and developing countries have dealt with budget cuts and cost-recovery requirements even before the growth of AI and the private weather industry, so further adoption of AI could exacerbate these trends.

For example, mobile weather apps have become the primary source of weather information for younger people \citep{Phan2018-ng}.  However, the apps' uniform presentation of weather information may result in users not being aware of more localized and rare weather hazards. A local government or television meteorologist on the other hand can tailor their communication to highlight these issues.

%%%%%%%%%%%%%%%%%%%%%%%%%%%%%%%%%%%%%%%%%%%%%%%%%%%%%%%%%%%%%%%%%%
\subsubsection{Lack of input or consent on data collection and model training}

As illustrated in the discussion of Figure \ref{fig:earthquake},  when collecting local data for training an AI model or when creating an AI model to inform decisions in or about a specific place, it is critical that the people affected and those with local expertise  are involved from the beginning \citep{Renn:1995,Stern:1996,Chilvers2009-fr,Voinov2016-yz,Pidgeon2021}. This is more typically known in the social sciences but not nearly as common in AI.

The types of data that are collected affect the AI models that can be applied and trained.  Understanding that data provides a source of power and that what is not collected cannot be used \citep{DataFeminism}, it is possible to see that AI could too easily be used to perpetuate environmental and climate injustices if data are not collected carefully.  

In environmental research it has long been noted that engaging those whose concerns the science is purportedly informing or addressing can improve models and outcomes, is a moral obligation, and is the virtuous course of action \citep[e.g.,][]{Lemos2005-tj}.  A core duty in all science is honesty, to respect the dignity and rights of persons \citep[e.g.,]{Keohane2014}.  In AI for environmental sciences this requires grappling with how to be transparent (i.e., show your work – \citep{DataFeminism}) about methodological biases and uncertainties that are often complex and poorly understood.  When scientific conditions such as sensitivity to training conditions or data provenance are not reported, that lack of transparency both disempowers and disrespects those the model may be intended to help, and can make a study irreplicable and even irreproducible \citep{NASEMReproduce2019}. 

It is thus critical to keep in mind that many types of data and knowledge already exist outside of the colonial framework in which most AI data has been collected.  For example, scientists must work directly with local leaders to incorporate Indigenous knowledge  \citep{EOSIndigenous2019,Hiwasaki2014-ji} in an ethical and responsible manner.

%%%%%%%%%%%%%%%%%%%%%%%%%%%%%%%%%%%%%%%%%%%%%%%%%%%%%%%%%%%%%%%%%%
\subsubsection{Scientists feeling disenfranchised}

There are two primary reasons why scientists might feel disenfranchised.

\noindent
\textbf{Scientists not fluent in AI might feel disenfranchised:}
Environmental scientists are absolutely essential for the development of meaningful AI tools for the their domains.  Currently, with a lack of education in AI for environmental sciences, leading scientists may feel excluded from  advancing state-of-the-art tools.  They must either learn about AI themselves or collaborate with AI scientists to be part of major future development in their field.  As a consequence they may feel less independent and less appreciated.
Clearly, we need more educational resources to help environmental scientists learn the basics of AI in order to empower them to play a leading role in future developments. 

%\noindent
\textbf{Required computational resources might limit research opportunities to privileged groups:}
While AI methods for many environmental science applications require only modest computational resources, some research now requires access to extensive high performance computing facilities, which limits this kind of research to those research groups privileged enough to have such access.  To level the playing field we as a community need to find ways to develop an open science model that provides access to computational resources also for other groups who need them \citep[e.g.,][]{Gentemann2021-nh}.

%%%%%%%%%%%%%%%%%%%%%%%%%%%%%%%%%%%%%%%%%%%%%%%%%%%%%%%%%%%%%%%%%%
\subsubsection{Increase of CO2 emissions due to computing}

It is well known that the increasing computational demands of AI training are responsible for a surprisingly large carbon footprint worldwide  \citep{xu2021survey,schwartz2020green}.  
Thus we need to consider, and limit, the carbon footprint from the computational needs of environmental science.   
Green deep learning (aka Green AI) is a new research field that appeals to researchers to pay attention to the carbon footprint of their research, and to focus on using lightweight and efficient methods \citep{xu2021survey}.
We should fully explore these new developments, including
%One under-explored research direction is to 
focusing on how to make ML models simpler for environmental science applications, which would have many advantages, including increased transparency and robustness \citep{Rudin2019}, but also lower computational needs and lower carbon footprints.

%%%%%%%%%%%%%%%%%%%%%%%%%%%%%%%%%%%%%%%%%%%%%%%%%%%%%%%%%%%%%%%%%%
\section{Discussion and Future work}

The scientific community is still grappling with the many ethical questions raised by the introduction of AI in general 
\citep{schmidt2021national,floridi2019establishing,WeaponsOfMath}
%dignum2019ensuring
and for earth science applications in particular
\citep{coeckelbergh2021ai,doorn2021artificial}.
The effort presented here represents the beginning of our work on developing a full understanding of the need for ethical, responsible, and trustworthy AI for the environmental sciences and of the interactions between ethical and responsible AI and trustworthy AI
%.  This is part of our work in 
within the {\it NSF AI Institute for Research on Trustworthy AI in Weather, Climate, and Coastal Oceanography (AI2ES, \url{ai2es.org})}.  In an upcoming paper, we plan to present guiding principles for ethical, responsible, and trustworthy AI for the environmental sciences.

%%%%%%%%%%%%%%%%%%%%%%%%%%%%%%%%%%%%%%%%%%%%%%%%%%%%%%%%%%%%%%%%%%
\begin{Backmatter}

\paragraph{Acknowledgments}
We are grateful for the graphic on radar coverage in the southeast US contributed by Jack Sillin (Cornell), the graphics contributed by Randy Chase (OU), and references contributed by Amy Burzynski (CSU).

\paragraph{Funding Statement}
This material is based upon work supported by the National Science Foundation under Grant No. ICER-2019758 and by the National Center for Atmospheric Research, which is a major facility sponsored by the National Science Foundation under Cooperative Agreement No. 1852977.

\paragraph{Competing Interests}
None.

\paragraph{Data Availability Statement}
This manuscript does not develop any new data or code.

\paragraph{Ethical Standards}
The research meets all ethical guidelines, including adherence to the legal requirements of the study country.

\paragraph{Author Contributions}
Conceptualization: All authors. Methodology: All authors. Data visualization: A.M and I.E. Writing original draft: A.M and I.E equally shared the majority of the writing of this paper. All authors contributed to the writing and approved the final submitted draft.

\paragraph{Supplementary Material}
None.
%State whether any supplementary material intended for publication has been provided with the submission.

\bibliographystyle{ametsoc2014}
\bibliography{clean_references}

\begin{thebibliography}{70}
\providecommand{\natexlab}[1]{#1}
\providecommand{\url}[1]{\texttt{#1}}
\renewcommand{\UrlFont}{\rmfamily}
\providecommand{\urlprefix}{URL }
\expandafter\ifx\csname urlstyle\endcsname\relax
  \providecommand{\doi}[1]{doi:\discretionary{}{}{}#1}\else
  \providecommand{\doi}{doi:\discretionary{}{}{}\begingroup
  \urlstyle{rm}\Url}\fi
\providecommand{\eprint}[2][]{\url{#2}}

\bibitem[{Adee(2020)}]{IEEEDeepFakes}
Adee, S., 2020: What are deepfakes and how are they created? \textit{{IEEE}
  Spectrum}, \urlprefix\url{https://spectrum.ieee.org/what-is-deepfake}.

\bibitem[{Allen and Tippett(2015)Allen, and Tippett}]{AllenHail2015}
Allen, J., and M.~Tippett, 2015: The characteristics of united states hail
  reports: 1955–2014. \textit{Electronic Journal of Severe Storms
  Meteorology}, \textbf{10}, 1--31.

\bibitem[{Barnes et~al.(2021)Barnes, Barnes,, and Gordillo}]{barnes2021adding}
Barnes, E.~A., R.~J. Barnes, and N.~Gordillo, 2021: Adding uncertainty to
  neural network regression tasks in the geosciences. \textit{arXiv preprint
  arXiv:2109.07250}.

\bibitem[{Benjamin(2019)}]{RaceAfterTechnology}
Benjamin, R., 2019: \textit{Race After Technology: Abolitionist Tools for the
  New Jim Code.} Cambridge: Polity.

\bibitem[{Cappucci(2020)}]{MpingWaPo}
Cappucci, M., 2020: {NOAA} storm-spotting app was suspended after being overrun
  with false and hateful reports. \textit{The Washington Post},
  \urlprefix\url{https://www.washingtonpost.com/weather/2020/07/14/noaa-app-mping-suspended/}.

\bibitem[{Cartier(2019)}]{EOSIndigenous2019}
Cartier, K. M.~S., 2019: Keeping indigenous science knowledge out of a colonial
  mold. \textit{{EOS} Science News by {AGU}},
  \urlprefix\url{https://eos.org/articles/keeping-indigenous-science-knowledge-out-of-a-colonial-mold}.

\bibitem[{Chase and McGovern(2022)Chase, and McGovern}]{chaseAMS2022}
Chase, R., and A.~McGovern, 2022: Deep learning parameter considerations when
  using radar and satellite measurements. \textit{21st Conference on Artificial
  Intelligence for Environmental Science at the 102th American Meteorological
  Society Annual Meeting}, AMS.

\bibitem[{Chase et~al.(2021)Chase, Nesbitt,, and McFarquhar}]{ChaseRadar2021}
Chase, R.~J., S.~W. Nesbitt, and G.~M. McFarquhar, 2021: A dual-frequency radar
  retrieval of two parameters of the snowfall particle size distribution using
  a neural network. \textit{Journal of Applied Meteorology and Climatology},
  \textbf{60~(3)}, 341 -- 359, \doi{10.1175/JAMC-D-20-0177.1},
  \urlprefix\url{https://journals.ametsoc.org/view/journals/apme/60/3/JAMC-D-20-0177.1.xml}.

\bibitem[{Chilson et~al.(2019)Chilson, Avery, McGovern, Bridge, Sheldon,, and
  Kelly}]{Chilson2019-uv}
Chilson, C., K.~Avery, A.~McGovern, E.~Bridge, D.~Sheldon, and J.~Kelly, 2019:
  Automated detection of bird roosts using {NEXRAD} radar data and
  convolutional neural networks. \textit{Remote Sens. Ecol. Conserv.},
  \textbf{5~(1)}, 20--32, \doi{10.1002/rse2.92},
  \urlprefix\url{https://onlinelibrary.wiley.com/doi/10.1002/rse2.92}.

\bibitem[{Chilvers(2009)}]{Chilvers2009-fr}
Chilvers, J., 2009: Deliberative and participatory approaches in environmental
  geography. \textit{A Companion to Environmental Geography}, Wiley-Blackwell,
  Oxford, UK, 400--417.

\bibitem[{Coeckelbergh(2021)}]{coeckelbergh2021ai}
Coeckelbergh, M., 2021: Ai for climate: freedom, justice, and other ethical and
  political challenges. \textit{AI and Ethics}, \textbf{1~(1)}, 67--72.

\bibitem[{Davila et~al.(2005)Davila, Marquart,, and Mullings}]{Davila2005-xk}
Davila, M., J.~W. Marquart, and J.~L. Mullings, 2005: beyond mother nature:
  contractor fraud in the wake of natural disasters. \textit{Deviant Behav.},
  \textbf{26~(3)}, 271--293, \doi{10.1080/01639620590927623},
  \urlprefix\url{https://doi.org/10.1080/01639620590927623}.

\bibitem[{deSouza and Kinney(2021)deSouza, and
  Kinney}]{desouza2021distribution}
deSouza, P., and P.~L. Kinney, 2021: On the distribution of low-cost pm 2.5
  sensors in the us: demographic and air quality associations. \textit{Journal
  of Exposure Science \& Environmental Epidemiology}, \textbf{31~(3)},
  514--524.

\bibitem[{{D'Ignazio} and Klein(2020){D'Ignazio}, and Klein}]{DataFeminism}
{D'Ignazio}, C., and L.~F. Klein, 2020: \textit{Data Feminism}. MIT Press, USA.

\bibitem[{Diochnos et~al.(2018)Diochnos, Mahloujifar,, and
  Mahmoody}]{ADV:Diochnosetal:neurips2018}
Diochnos, D., S.~Mahloujifar, and M.~Mahmoody, 2018: Adversarial risk and
  robustness: General definitions and implications for the uniform
  distribution. \textit{Advances in Neural Information Processing Systems 31},
  S.~Bengio, H.~Wallach, H.~Larochelle, K.~Grauman, N.~Cesa-Bianchi, and
  R.~Garnett, Eds., Curran Associates, Inc., 10\,359--10\,368,
  \urlprefix\url{http://papers.nips.cc/paper/8237-adversarial-risk-and-robustness-general-definitions-and-implications-for-the-uniform-distribution.pdf}.

\bibitem[{Doorn(2021)}]{doorn2021artificial}
Doorn, N., 2021: Artificial intelligence in the water domain: Opportunities for
  responsible use. \textit{Science of the Total Environment}, \textbf{755},
  142\,561.

\bibitem[{Duporge et~al.(2021)Duporge, Isupova, Reece, Macdonald,, and
  Wang}]{Duporge2021-cb}
Duporge, I., O.~Isupova, S.~Reece, D.~W. Macdonald, and T.~Wang, 2021: Using
  very‐high‐resolution satellite imagery and deep learning to detect and
  count african elephants in heterogeneous landscapes. \textit{Remote Sens.
  Ecol. Conserv.}, \textbf{7~(3)}, 369--381, \doi{10.1002/rse2.195},
  \urlprefix\url{https://onlinelibrary.wiley.com/doi/10.1002/rse2.195}.

\bibitem[{Ebert-Uphoff and Hilburn(2020)Ebert-Uphoff, and
  Hilburn}]{ebert2020evaluation}
Ebert-Uphoff, I., and K.~Hilburn, 2020: Evaluation, tuning, and interpretation
  of neural networks for working with images in meteorological applications.
  \textit{Bulletin of the American Meteorological Society}, \textbf{101~(12)},
  E2149--E2170.

\bibitem[{Edwards et~al.(2018)Edwards, Allen,, and
  Carbin}]{EdwardsWindReliability:2018}
Edwards, R., J.~T. Allen, and G.~W. Carbin, 2018: Reliability and
  climatological impacts of convective wind estimations. \textit{Journal of
  Applied Meteorology and Climatology}, \textbf{57~(8)}, 1825 -- 1845,
  \doi{10.1175/JAMC-D-17-0306.1},
  \urlprefix\url{https://journals.ametsoc.org/view/journals/apme/57/8/jamc-d-17-0306.1.xml}.

\bibitem[{Fink(2019)}]{fink2019high}
Fink, S., 2019: This high-tech solution to disaster response may be too good to
  be true. \textit{New York Times. Retrieved August}, \textbf{12}, 2020,
  \urlprefix\url{https://www.nytimes.com/2019/08/09/us/emergency-response-disaster-technology.html}.

\bibitem[{Floridi(2019)}]{floridi2019establishing}
Floridi, L., 2019: Establishing the rules for building trustworthy ai.
  \textit{Nature Machine Intelligence}, \textbf{1~(6)}, 261--262.

\bibitem[{Gagne et~al.(2020)Gagne, Christensen,, and {others}}]{Gagne2020-kj}
Gagne, D.~J., H.~M. Christensen, and {others}, 2020: Machine learning for
  stochastic parameterization: Generative adversarial networks in the lorenz'96
  model. \textit{Journal of Advances},
  \urlprefix\url{https://agupubs.onlinelibrary.wiley.com/doi/abs/10.1029/2019MS001896}.

\bibitem[{Gagne et~al.(2017)Gagne, McGovern, Haupt,, and
  Williams}]{Gagne2017-vw}
Gagne, D.~J., A.~McGovern, S.~E. Haupt, and J.~K. Williams, 2017: Evaluation of
  statistical learning configurations for gridded solar irradiance forecasting.
  \textit{Solar Energy}, \textbf{150}, 383--393,
  \doi{10.1016/j.solener.2017.04.031},
  \urlprefix\url{https://www.sciencedirect.com/science/article/pii/S0038092X17303158}.

\bibitem[{Geiss and Hardin(2021)Geiss, and Hardin}]{Geiss2021-ob}
Geiss, A., and J.~C. Hardin, 2021: Inpainting radar missing data regions with
  deep learning. \textit{Atmos. Meas. Tech.}, \textbf{14~(12)}, 7729--7747,
  \doi{10.5194/amt-14-7729-2021},
  \urlprefix\url{https://amt.copernicus.org/articles/14/7729/2021/}.

\bibitem[{Gensini et~al.(2021)Gensini, Walker, Ashley,, and
  Taszarek}]{Gensini2021}
Gensini, V.~A., C.~C. Walker, S.~Ashley, and M.~Taszarek, 2021: Machine
  learning classification of significant tornadoes and hail in the united
  states using era5 proximity soundings. \textit{Weather and Forecasting},
  \textbf{36~(6)}, 2143 -- 2160, \doi{10.1175/WAF-D-21-0056.1},
  \urlprefix\url{https://journals.ametsoc.org/view/journals/wefo/36/6/WAF-D-21-0056.1.xml}.

\bibitem[{Gentemann et~al.(2021)Gentemann, Holdgraf, Abernathey, Crichton,
  Colliander, Kearns, Panda,, and Signell}]{Gentemann2021-nh}
Gentemann, C.~L., C.~Holdgraf, R.~Abernathey, D.~Crichton, J.~Colliander, E.~J.
  Kearns, Y.~Panda, and R.~P. Signell, 2021: Science storms the cloud.
  \textit{AGU Advances}, \textbf{2~(2)}.

\bibitem[{Goodfellow et~al.(2014)Goodfellow, Pouget-Abadie, Mirza, Xu,
  Warde-Farley, Ozair, Courville,, and Bengio}]{goodfellow2014generative}
Goodfellow, I., J.~Pouget-Abadie, M.~Mirza, B.~Xu, D.~Warde-Farley, S.~Ozair,
  A.~Courville, and Y.~Bengio, 2014: Generative adversarial nets.
  \textit{Advances in neural information processing systems}, \textbf{27}.

\bibitem[{Goodfellow et~al.(2018)Goodfellow, McDaniel,, and
  Papernot}]{GoodfellowEtAl:MakeMLRobust}
Goodfellow, I.~J., P.~D. McDaniel, and N.~Papernot, 2018: {Making machine
  learning robust against adversarial inputs}. \textit{Communications of the
  {ACM}}, \textbf{61~(7)}, 56--66.

\bibitem[{Hilburn et~al.(2021)Hilburn, Ebert-Uphoff,, and
  Miller}]{hilburn2021development}
Hilburn, K.~A., I.~Ebert-Uphoff, and S.~D. Miller, 2021: Development and
  interpretation of a neural-network-based synthetic radar reflectivity
  estimator using {GOES-R} satellite observations. \textit{Journal of Applied
  Meteorology and Climatology}, \textbf{60~(1)}, 3--21.

\bibitem[{Hill and Schumacher(2021)Hill, and Schumacher}]{Hill2021}
Hill, A.~J., and R.~S. Schumacher, 2021: Forecasting excessive rainfall with
  random forests and a deterministic convection-allowing model. \textit{Weather
  and Forecasting}, \textbf{36~(5)}, 1693 -- 1711,
  \doi{10.1175/WAF-D-21-0026.1},
  \urlprefix\url{https://journals.ametsoc.org/view/journals/wefo/36/5/WAF-D-21-0026.1.xml}.

\bibitem[{Hiwasaki et~al.(2014)Hiwasaki, Luna, {Syamsidik},, and
  Shaw}]{Hiwasaki2014-ji}
Hiwasaki, L., E.~Luna, {Syamsidik}, and R.~Shaw, 2014: Process for integrating
  local and indigenous knowledge with science for hydro-meteorological disaster
  risk reduction and climate change adaptation in coastal and small island
  communities. \textit{International Journal of Disaster Risk Reduction},
  \textbf{10}, 15--27.

\bibitem[{Karpatne et~al.(2016)Karpatne, Jiang, Vatsavai, Shekhar,, and
  Kumar}]{Karpatne2016-ti}
Karpatne, A., Z.~Jiang, R.~R. Vatsavai, S.~Shekhar, and V.~Kumar, 2016:
  Monitoring {Land-Cover} changes: A {Machine-Learning} perspective.
  \textit{IEEE Geoscience and Remote Sensing Magazine}, \textbf{4~(2)}, 8--21,
  \doi{10.1109/MGRS.2016.2528038},
  \urlprefix\url{http://dx.doi.org/10.1109/MGRS.2016.2528038}.

\bibitem[{Keohane et~al.(2014)Keohane, Lane,, and Oppenheimer}]{Keohane2014}
Keohane, R.~O., M.~Lane, and M.~Oppenheimer, 2014: The ethics of scientific
  communication under uncertainty. \textit{Politics, Philosophy \& Economics},
  \textbf{13~(4)}, 343--368, \doi{10.1177/1470594X14538570},
  \urlprefix\url{https://doi.org/10.1177/1470594X14538570},
  \eprint{https://doi.org/10.1177/1470594X14538570}.

\bibitem[{Lagerquist et~al.(2020)Lagerquist, McGovern, Homeyer, II,, and
  Smith}]{Lagerquist2020}
Lagerquist, R., A.~McGovern, C.~R. Homeyer, D.~J.~G. II, and T.~Smith, 2020:
  Deep learning on three-dimensional multiscale data for next-hour tornado
  prediction. \textit{Monthly Weather Review}, \textbf{148~(7)}, 2837 -- 2861,
  \doi{10.1175/MWR-D-19-0372.1},
  \urlprefix\url{https://journals.ametsoc.org/view/journals/mwre/148/7/mwrD190372.xml}.

\bibitem[{Lagerquist et~al.(2021)Lagerquist, Stewart, Ebert-Uphoff,, and
  ChristinaKumler}]{Lagerquist2021}
Lagerquist, R., J.~Q. Stewart, I.~Ebert-Uphoff, and ChristinaKumler, 2021:
  Using deep learning to nowcast the spatial coverage of convection from
  himawari-8 satellite data. \textit{Monthly Weather Review},
  \textbf{149~(12)}, 3897 -- 3921, \doi{10.1175/MWR-D-21-0096.1},
  \urlprefix\url{https://journals.ametsoc.org/view/journals/mwre/149/12/MWR-D-21-0096.1.xml}.

\bibitem[{Lemos and Morehouse(2005)Lemos, and Morehouse}]{Lemos2005-tj}
Lemos, M.~C., and B.~J. Morehouse, 2005: The co-production of science and
  policy in integrated climate assessments. \textit{Glob. Environ. Change},
  \textbf{15~(1)}, 57--68.

\bibitem[{Lin et~al.(2019)}]{Lin2019-xd}
Lin, T.-Y., and Coauthors, 2019: {M} ist {N} et : Measuring historical bird
  migration in the {US} using archived weather radar data and convolutional
  neural networks. \textit{Methods Ecol. Evol.}, \textbf{10~(11)}, 1908--1922,
  \doi{10.1111/2041-210x.13280},
  \urlprefix\url{https://onlinelibrary.wiley.com/doi/10.1111/2041-210X.13280}.

\bibitem[{Matias(2021)}]{GoogleFloodBlog2021}
Matias, Y., 2021: Expanding our ml-based flood forecasting.
  \urlprefix\url{https://blog.google/technology/ai/expanding-our-ml-based-flood-forecasting/}.

\bibitem[{McGovern et~al.(2019)McGovern, Lagerquist, Gagne, Jergensen, Elmore,
  Homeyer,, and Smith}]{McGovern2019_bams}
McGovern, A., R.~Lagerquist, D.~Gagne, G.~Jergensen, K.~Elmore, C.~Homeyer, and
  T.~Smith, 2019: {Making the black box more transparent: Understanding the
  physical implications of machine learning}. \textit{Bulletin of the American
  Meteorological Society}, \textbf{100~(11)}, 2175--2199,
  \doi{10.1175/BAMS-D-18-0195.1}.

\bibitem[{Mesonet(2020)}]{MesonetTop20}
Mesonet, O., 2020: Top 20 extreme weather events in mesonet history.
  \urlprefix\url{https://www.mesonet.org/20th/}.

\bibitem[{Mithal et~al.(2018)Mithal, Nayak, Khandelwal, Kumar, Nemani,, and
  Oza}]{Mithal2018-gp}
Mithal, V., G.~Nayak, A.~Khandelwal, V.~Kumar, R.~Nemani, and N.~C. Oza, 2018:
  Mapping burned areas in tropical forests using a novel machine learning
  framework. \textit{Remote Sensing}, \textbf{10~(1)}, 69,
  \doi{10.3390/rs10010069},
  \urlprefix\url{https://www.mdpi.com/2072-4292/10/1/69}.

\bibitem[{Molnar(2018)}]{Molnar2018}
Molnar, C., 2018: \textit{{Interpretable Machine Learning: A Guide for Making
  Black Box Models Explainable}}. {Leanpub},
  \urlprefix\url{https://christophm.github.io/interpretable-ml-book/}.

\bibitem[{Murillo and Homeyer(2019)Murillo, and Homeyer}]{MurilloMesh2019}
Murillo, E.~M., and C.~R. Homeyer, 2019: Severe hail fall and hailstorm
  detection using remote sensing observations. \textit{Journal of Applied
  Meteorology and Climatology}, \textbf{58~(5)}, 947 -- 970,
  \doi{10.1175/JAMC-D-18-0247.1},
  \urlprefix\url{https://journals.ametsoc.org/view/journals/apme/58/5/jamc-d-18-0247.1.xml}.

\bibitem[{{National Academies of Sciences, Engineering, and Medicine}
  et~al.(2019)}]{NASEMReproduce2019}
{National Academies of Sciences, Engineering, and Medicine}, and Coauthors,
  2019: \textit{Reproducibility and Replicability in Science}. National
  Academies Press.

\bibitem[{Nelson et~al.(2010)Nelson, Rubinstein, Huang, Joseph,, and
  Tygar}]{Adversarial:models}
Nelson, B., B.~I.~P. Rubinstein, L.~Huang, A.~D. Joseph, and J.~D. Tygar, 2010:
  {Classifier Evasion: Models and Open Problems}. \textit{PSDM}, 92--98.

\bibitem[{{O'Neil}(2016)}]{WeaponsOfMath}
{O'Neil}, C., 2016: \textit{Weapons of Math Destruction: How Big Data Increases
  Inequality and Threatens Democracy}. Crown Publishing Group, USA.

\bibitem[{Orescanin et~al.(2021)Orescanin, Petkovi{\'c}, Powell, Marsh,, and
  Heslin}]{orescanin2021bayesian}
Orescanin, M., V.~Petkovi{\'c}, S.~W. Powell, B.~R. Marsh, and S.~C. Heslin,
  2021: Bayesian deep learning for passive microwave precipitation type
  detection. \textit{IEEE Geoscience and Remote Sensing Letters}.

\bibitem[{Phan et~al.(2018)Phan, Montz, Curtis,, and Rickenbach}]{Phan2018-ng}
Phan, M.~D., B.~E. Montz, S.~Curtis, and T.~M. Rickenbach, 2018: Weather on the
  go: An assessment of smartphone mobile weather application use among college
  students. \textit{Bull. Am. Meteorol. Soc.}, \textbf{99~(11)}, 2245--2257,
  \doi{10.1175/BAMS-D-18-0020.1},
  \urlprefix\url{https://journals.ametsoc.org/view/journals/bams/99/11/bams-d-18-0020.1.xml?tab_body=pdf}.

\bibitem[{Pidgeon(2021)}]{Pidgeon2021}
Pidgeon, N., 2021: Engaging publics about environmental and technology risks:
  frames, values and deliberation. \textit{Journal of Risk Research},
  \textbf{24~(1)}, 28--46, \doi{10.1080/13669877.2020.1749118},
  \urlprefix\url{https://doi.org/10.1080/13669877.2020.1749118}.

\bibitem[{Potvin et~al.(2019)Potvin, Broyles, Skinner, Brooks,, and
  Rasmussen}]{Potvin2019tornadoreports}
Potvin, C.~K., C.~Broyles, P.~S. Skinner, H.~E. Brooks, and E.~Rasmussen, 2019:
  A bayesian hierarchical modeling framework for correcting reporting bias in
  the u.s. tornado database. \textit{Weather and Forecasting}, \textbf{34},
  15--30, \doi{10.1175/WAF-D-18-0137.1}.

\bibitem[{Ras et~al.(2018)Ras, van Gerven,, and Haselager}]{ras2018explanation}
Ras, G., M.~van Gerven, and P.~Haselager, 2018: Explanation methods in deep
  learning: Users, values, concerns and challenges. \textit{Explainable and
  Interpretable Models in Computer Vision and Machine Learning}, Springer,
  19--36.

\bibitem[{Ravuri et~al.(2021)}]{Ravuri2021-dg}
Ravuri, S., and Coauthors, 2021: Skilful precipitation nowcasting using deep
  generative models of radar. \textit{Nature}, \textbf{597~(7878)}, 672--677.

\bibitem[{{RE}(2018)}]{InsuredLosses2018}
{RE}, S., 2018: Preliminary sigma estimates for 2018: global insured losses of
  usd 79 billion are fourth highest on sigma records.
  \urlprefix\url{https://www.swissre.com/media/news-releases/nr_20181218_sigma_estimates_for_2018.html}.

\bibitem[{Reichstein et~al.(2019)Reichstein, Camps-Valls, Stevens, Jung,
  Denzler, Carvalhais,, and Prabhat}]{Reichstein2019}
Reichstein, M., G.~Camps-Valls, B.~Stevens, M.~Jung, J.~Denzler, N.~Carvalhais,
  and Prabhat, 2019: {Deep learning and process understanding for data-driven
  Earth system science}. \textit{Nature}, \textbf{566}, 195--204,
  \urlprefix\url{https://doi.org/10.1038/s41586-019-0912-1}.

\bibitem[{Renn et~al.(1995)Renn, Webler,, and Wiedemann}]{Renn:1995}
Renn, O., T.~Webler, and P.~Wiedemann, 1995: \textit{Fairness and Competence in
  Citizen Participation: Evaluating Models for Environmental Discourse}. Kluwer
  Academic Press, Dordrecht.

\bibitem[{Rudin(2019)}]{Rudin2019}
Rudin, C., 2019: Stop explaining black box machine learning models for high
  stakes decisions and use interpretable models instead. \textit{Nat Mach
  Intell}, \textbf{1}, 206–215,
  \urlprefix\url{https://doi.org/10.1038/s42256-019-0048-x}.

\bibitem[{Samek et~al.(2019)Samek, Montavon, Vedaldi, Hansen,, and
  Muller}]{XAI_book}
Samek, W., G.~Montavon, A.~Vedaldi, L.~Hansen, and K.-R. Muller, Eds., 2019:
  \textit{Explainable {AI}: Interpreting, Explaining and Visualizing Deep
  Learning}. Springer Nature.

\bibitem[{Schmidt et~al.(2021)}]{schmidt2021national}
Schmidt, E., and Coauthors, 2021: National security commission on artificial
  intelligence (ai). Tech. rep., National Security Commission on Artificial
  Intellegence.

\bibitem[{Schumacher et~al.(2021)Schumacher, Hill, Klein, Nelson, Erickson,
  Trojniak,, and Herman}]{Schumacher2021}
Schumacher, R.~S., A.~J. Hill, M.~Klein, J.~A. Nelson, M.~J. Erickson, S.~M.
  Trojniak, and G.~R. Herman, 2021: From random forests to flood forecasts: A
  research to operations success story. \textit{Bulletin of the American
  Meteorological Society}, \textbf{102~(9)}, E1742 -- E1755,
  \doi{10.1175/BAMS-D-20-0186.1},
  \urlprefix\url{https://journals.ametsoc.org/view/journals/bams/102/9/BAMS-D-20-0186.1.xml}.

\bibitem[{Schwartz et~al.(2020)Schwartz, Dodge, Smith,, and
  Etzioni}]{schwartz2020green}
Schwartz, R., J.~Dodge, N.~A. Smith, and O.~Etzioni, 2020: Green ai.
  \textit{Communications of the ACM}, \textbf{63~(12)}, 54--63.

\bibitem[{Scotti(2019)}]{WorldEconomicInsurance}
Scotti, V., 2019: How cities can become more resilient to climate change.
  \textit{World Economic Forum},
  \urlprefix\url{https://www.weforum.org/agenda/2019/06/can-we-be-resilient-to-climate-change/}.

\bibitem[{Shepherd(2021)}]{ShepherdRadar2021}
Shepherd, M., 2021: Are black and rural residents in the south more vulnerable
  to tornadoes due to radar gaps? \textit{Forbes},
  \urlprefix\url{https://www.forbes.com/sites/marshallshepherd/2021/03/20/are-black-and-rural-residents-in-the-south-more-vulnerable-to-tornadoes-due-to-radar-gaps/?sh=58c1d4014988}.

\bibitem[{Sillin(2021)}]{Sillin2021}
Sillin, J., 2021: Twitter post.
  \urlprefix\url{https://twitter.com/JackSillin/status/1372957704138981378?s=20}.

\bibitem[{Sorge(2018)}]{InsuranceFraud}
Sorge, G., 2018: Weather-related events and insurance fraud. \textit{Property
  Casualty 360},
  \urlprefix\url{https://www.propertycasualty360.com/2018/10/02/weather-related-events-and-insurance-fraud/?slreturn=20211026223239}.

\bibitem[{Staff(2021)}]{HurricaneOutages}
Staff, W., 2021: Recovery time unclear, check map for your parish.
  \textit{WWLTV},
  \urlprefix\url{https://www.wwltv.com/article/weather/hurricane/widespread-power-outages-reported-9400-in-the-dark/289-d8a78748-9a37-4937-90af-0d2d7cb3fbd6}.

\bibitem[{Stern and Fineberg.(1996)Stern, and Fineberg.}]{Stern:1996}
Stern, P.~C., and H.~C. Fineberg., 1996: \textit{Understanding Risk: Informing
  Decisions in a Democratic Society}. US National Research Council, Washington
  DC.

\bibitem[{Tan(2020)}]{tan2020film}
Tan, C., 2020: Film: Coded bias-slave to the algorithm. \textit{Big Issue
  Australia}, \textbf{~(617)}, 30--31.

\bibitem[{Vaccari and Chadwick(2020)Vaccari, and Chadwick}]{Vaccari2020-jh}
Vaccari, C., and A.~Chadwick, 2020: Deepfakes and disinformation: Exploring the
  impact of synthetic political video on deception, uncertainty, and trust in
  news. \textit{Social Media + Society}, \textbf{6~(1)}, 2056305120903\,408.

\bibitem[{Voinov et~al.(2016)Voinov, Kolagani, McCall, Glynn, Kragt, Ostermann,
  Pierce,, and Ramu}]{Voinov2016-yz}
Voinov, A., N.~Kolagani, M.~K. McCall, P.~D. Glynn, M.~E. Kragt, F.~O.
  Ostermann, S.~A. Pierce, and P.~Ramu, 2016: Modelling with stakeholders --
  next generation. \textit{Environmental Modelling \& Software}, \textbf{77},
  196--220.

\bibitem[{Xu et~al.(2021)Xu, Zhou, Fu, Zhou,, and Li}]{xu2021survey}
Xu, J., W.~Zhou, Z.~Fu, H.~Zhou, and L.~Li, 2021: A survey on green deep
  learning. \textit{arXiv preprint arXiv:2111.05193}.

\end{thebibliography}

\end{Backmatter}

\end{document}